# Novel measures based on the Kolmogorov complexity for use in complex system behavior studies and time series analysis

D.T. Mihailović[1], G. Mimić[2], Nikolić-Đorić[3] and I. Arsenić[1]

[1] Faculty of Agriculture, Division of Meteorology and Biophysics, University of Novi Sad, Dositeja Obradovica Sq. 8, 21000 Novi Sad, Serbia
[2] Faculty of Sciences, Department of Physics, University of Novi Sad, Dositeja Obradovica Sq. 3, 21000 Novi Sad, Serbia
[3] Faculty of Agriculture, Division of Statistics, University of Novi Sad, Dositeja Obradovica Sq. 8, 21000 Novi Sad, Serbia

**Abstract**

We have proposed novel measures based on the Kolmogorov complexity for use in complex system behavior studies and time series analysis. We have considered background of the Kolmogorov complexity and also we have discussed meaning of the physical as well as other complexities. To get better insights into the complexity of complex systems and time series analysis we have introduced the three novel measures based on the Kolmogorov complexity: (i) the Kolmogorov complexity spectrum, (ii) the Kolmogorov complexity spectrum highest value and (iii) the overall Kolmogorov complexity. The characteristics of these measures have been tested using a generalized logistic equation. Finally, the proposed measures have been applied on different time series originating from: the model output (the biochemical substance exchange in a multi-cell system), four different geophysical phenomena (dynamics of: river flow, long term precipitation, indoor $^{222}$Rn concentration and UV radiation dose) and economy (stock prices dynamics). Results which are obtained offer deeper insights into complexity of the system dynamics behavior and time series analysis when the proposed complexity measures are applied.

**Keywords**: Complexity • Physical complexity • Kolmogorov complexity spectrum • Kolmogorov complexity spectrum highest value • Overall Kolmogorov complexity • Time series

## 1. Introduction

The issue of complexity has been touching the scientific community intesively during the last three decades. This is happening on the epistemological as well as the methodological level. First, here we shortly outline the currently achieved epistemological level about this subject. The complexity is one of the aspects of self-organisation whose fundamental quality is an emergence. Self-organizing systems are complex systems. As it concisely elaborated in [1] the term *complexity* has three levels of meaning: (1) there is self-organization and emergence in complex systems [2], (2) complex systems are not organized centrally, but in a

[1] Corresponding author (D.T. Mihailović)

*Telephone*: +381216350552

*E-mail address*: guto@polj.uns.ac.rs

distributed manner; there are many connections between the system's parts [2,3], (3) it is difficult to model complex systems and to predict their behavior even if one knows to a large extent the parts of such systems and the connections between the parts [2,4]. The complexity of a system depends on the number of its elements and connections between the elements (the system´s structure). In the review paper Crutchfield [5] has underlined: „ Spontaneous organization, as a common phenomenon, reminds us of a more basic, nagging puzzle. If, as Poincaré found, chaos is endemic to dynamics, why is the world not a mass of randomness? The world is, in fact, quite structured, and we now know several of the mechanisms that shape microscopic fluctuations as they are amplified to macroscopic patterns. Critical phenomena in statistical mechanics [6] and pattern formation in dynamics [7,8] are two arenas that explain in predictive detail how spontaneous organization works. Moreover, everyday experience shows us that nature inherently organizes; it generates pattern. Pattern is as much the fabric of life as life's unpredictability". These sentences are also related to the phenomenon of the complexity of systems in many disciplines, ranging from philosophy and cognitive science to evolutionary and developmental biology and particle astrophysics [5,9 and refrences herein].

There exist a lot of complexity measures in complex system behavior and time series analysis. In particular, in the focus is analysis of time series since the only available evidence about the nature of complex system come through a time series. According to Zunino [10 and references herein], the recorded signals from experimental measurements give us useful information to establish the deterministic or stochastic character of the system under analysis, but the task to discern between regular, chaotic, and stochastic dynamics from complex time series is a critical issue. In the study of complex system behavior and time series analysis, an important part is played by symbolic sequences, since it is believed that most systems whose complexity we intent to estimate can be reduced to them [11]. Thus, in searching for an adequate measure for the complexity (for example physical) of sequences, it is difficult to do that consistently. In particular, measures of complexity that are based on the Kolmogorov complexity [12], useful in signal analysis, are measures of randomness rather than complexity. This complexity is not able to discern between signals with different amplitude variations and similar random components.

The purpose of this paper is to introduce novel complexity measures based on the Kolmogorov complexity for use in complex systems behavior and time series analysis, which can offer deeper insights into these issues. Note that in this paper we will deal specifically with the physical complexity. However, without loosing the generality the methods proposed can be applied as complexity measures for other kind of complexity. We do that through the following steps. In Section 2 we consider in what extent Kolmogorov complexity enlightens the physical complexity. The proposed measures based on the Kolmogorov complexity (the Kolmogorov complexity spectrum, the Kolmogorov complexity spectrum highest value and the overall Kolmogorov complexity), we describe in Section 3. In Section 4 we apply the proposed measures on different time series originating from: the model output (the biochemical substance exchange in a multi-cell system), four different geophysical phenomena dynamics (river flow rate, long term precipitation amount, indoor $^{222}$Rn concentration, UV radiation dose) and economy (stock prices dynamics). Concluding remarks are summarized in Section 5.

**2. In what extent Kolmogorov complexity enlightens the physical complexity?**

In this section, through several steps, we consider the Kolmogorov complexity, which is seen through its applicability in illuminating the physical complexity.

*2.1 Kolmogorov complexity*

The Kolmogorov complexity $K(x)$ of an object $x$ is the length, in bits, of the smallest program (in bits) that when run on a Universal Turing Machine ($U$) outputs $K(x)$ and then stops with the execution. This complexity is maximized for random strings. Thus, $K(x) = |\text{Print}(x)|$. That is, the shortest program to get a $U$ to produce is to just hand the computer a copy of and say "print this" [13]. This measure was independently developed by Andrey N. Kolmogorov in the late 1960s [14]. A good introduction to the Kolmogorov complexity (in further text KL) can be found in [12] and with a comprehensive description in [15]. On the basis of Kolmogorov's idea, Lempel and Ziv [16] developed an algorithm (LZA), which is often used in assessing the randomness of finite sequences as a measure of its disorder.

The Kolmogorov complexity of a time series $\{x_i\}$, $i = 1,2,3,4,...,N$ by the LZA algorithm can be summarized as follows. *Step A:* Encode the time series by constructing a sequence $S$ consisting of the characters 0 and 1 written as $\{s(i)\}$, $i=1,2,3,4,…,N$, according to the rule

$$s(i) = \begin{cases} 0 & x_i < x_* \\ 1 & x_i \geq x_* \end{cases}. \qquad (1)$$

Here $x_*$ is a threshold that should be properly chosen. The mean value of the time series has often been used as the threshold [17]. Depending on the application, other encoding schemes are also available [18]. *Step B:* Calculate the complexity counter $C(N)$, which is defined as the minimum number of distinct patterns contained in a given character sequence [19]; $c(N)$ is a function of the length of the sequence $N$. The value of $c(N)$ is approaching an ultimate value $b(N)$ as $N$ approaching infinite, i.e.

$$c(N) = O(b(N)), \quad b(N) = \frac{N}{\log_2 N}. \qquad (2)$$

*Step 3*: Calculate the normalized complexity measure $C_k(N)$, which is defined as

$$C_k(N) = \frac{c(N)}{b(N)} = c(N) \frac{\log_2 N}{N}. \qquad (3)$$

The $C_k(N)$ is a parameter to represent the information quantity contained in a time series, and it is to be a 0 for a periodic or regular time series and to be a 1 for a random time series, if $N$ is large enough. For a non-linear time series, $C_k(N)$ is to be between 0 and 1. Let us note that Hu et al. [20] derived analytic expression for $C_k$ in the Kolmogorov complexity, for regular and random sequences. In addition they showed that for the shorter length of the time series, the larger $C_k$ value and correspondingly the complexity for a random sequence can be considerably larger than 1.

*2.2 Kolmogorov complexity versus physical complexity*

*2.2.1 Preliminaries*

The term physical system covers many different things. In this paper the word system we use in, has a technical meaning, i.e., as the portion of the physical universe, which is chosen for analysis. Everything outside the system is known as the environment which in analysis is ignored except for its effects on the system. Generally, in order to simplify the analysis as much as possible we cut the links between system and the world, what is a free choice. Thus, an isolated system is an example for which is supposed to have negligible interaction with its environment. Often a system viewed in this sense is chosen to correspond to the more usual meaning of system like as a particular machine. Let us note that physical systems are often more esoteric like, for example, an atom, the water in a lake, or indeed the water in the left-hand half of a lake. All of them can be considered as physical systems. Also, when we study the loss of coherence or ordering of the phase angles between the components of a system in a quantum superposition (quantum decoherence - when a system interacts with its environment in a thermodynamically irreversible way [21]), the system may refer to the macroscopic properties of an object (e.g. the position of a pendulum bob), while the relevant environment may be the internal degrees of freedom, described classically by the pendulum's thermal vibrations. However, regardless of which physical system, or any other complex system, we consider the term complexity is always present either explicitly or implicitly

*2.2.2 Physical complexity*

When we use the term *complexity* in physical systems we explicitly think that it is a measure of the probability of the state vector of the system. It is a mathematical measure, one in which two distinct states are never be combined into a composite whole and considered equal, as is done for the notion of entropy in statistical mechanics. Therefore, this term should not be equalized with *entropy* in statistical mechanics. According to Adami [22] the physical complexity of a sequence "refers to the amount of information that is stored in that sequence about a particular environment". This should not be confused with mathematical (Kolmogorov) complexity; it is a distinct mathematical complexity, which only deals either with the intrinsic regularity or irregularity of a sequence in this case. Namely, for any two strings $x, y \in P^*$ ($P^*$ is the set of all finite binary strings), the Kolmogorov comlexity of given $x$ is $K(x|y) = min_p \{|p| : U(p, y) = x\}$ where $U(p, y)$ is the output of the program $p$ with auxiliary input $y$ when it is run in the machine $U$. For any time constructible $t$, the $t$-time-bounded Kolmogorov complexity of $x$ given $y$ is, $K^t(x|y) = min_p \{|p| : U(p, y) = x \text{ in at most } (t|p|) \text{ steps}\}$. Let us note that the regularity of a sequence we are talking about is just a reflection of the unchanging laws of mathematics. It is not a reflection of the physical world where such a sequence may mean something. The Kolmogorov complexity does not measure pattern or structure or correlation or organization. Structure or pattern is maximized for neither high nor low randomness. If we follow Grassberger [23], on the level of intuition it can be accepted as something that is placed in between uniformity and total randomness (Fig. 1). Let us note that the structural complexity versus randomness relation is just one of many possible behaviors. Different systems have different structural complexity versus randomness plots [13]. There is no "universal" complexity relationship, which is clearly established in the scientific literature.

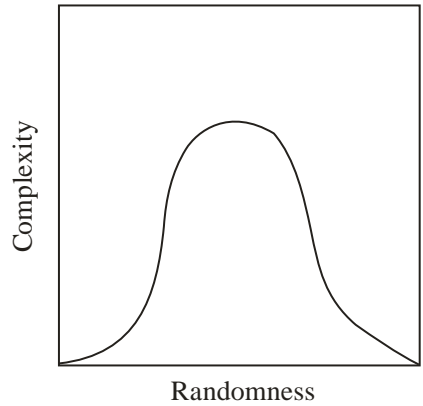

Fig. 1 Complexity versus randomness plotted following the physical intuition [13,23].

Adami and Cerf [11], proposing a measure of physical complexity, cleverly observed that it should closely correspond to our intuition. In addition they enhanced that it can consistently be defined within information theory. In studying the complex systems, an important step in using this measure is connected with symbolic sequences. Namely, it is believed that most systems whose complexity we would like to estimate can be reduced to them. The idea that the regularity of such a created string is in any way connected to its complexity (as in Kolmogorov theory), applied on different systems (physical, biological, chemical, social, literature etc.), in the absence of their environment within which the string is to be interpreted, thoroughly meaningless. There is no doubt that it is possible to establish a coding system, for example, such that entire "The Tin Drum" by Grass [24] is represented in terms of a uniform (and thus "regular") string of vanishing the Kolmogorov complexity. Thus, one possible coding system could be invented in the following way: to event it is assigned 1 (one), when Oskar Makowski strikes the drum when it is described explicitly. Otherwise, it is 0 (zero). Although, this event is presented metaphorically on other events in the book, evidently, in such a case the literature complexity of the string is hidden in the coding rules which relate the string to the ensemble of books as its environment. Thus, the complexity of a string representing the physical (or any other) complexity can be determined only by analyzing its correlation with a physical or corresponding environment.

## 3. Novel measures based on the Kolmogorov complexity

The quantification of the complexity of a system is one of the aims of non-linear time series analysis. Complexity of the system is hidden in the dynamics of the system. However, if there is no recognizable structure in the system, it is considered to be stochastic. Because of the occurrence of noise, spurious experimental result and artifacts in various forms, it is often not easy to get reliable information from a series of measurements. The time series of some physical quantity is only information about physical state, we can obtain by either measurement or from a modeling process. Therefore, acccordingly that is only source for establishing the level of physical complexity. The exact states of an observed physical system are translated into a sequence of symbols via a measurement channel. This process is described by a parameterized partition $M_\varepsilon$, of the state space, consisting of cells of size $\varepsilon$ that are sampled every $\tau$ time units. A measurement sequence consists of the successive elements of $M_\varepsilon$, visited over time by the system's state. Using the instrument $\{M_\varepsilon, \tau\}$ we get information as a sequence of states $\{x_i\}$. Here, we consider a possible way how to calculate the physical complexity of the system, i.e. complexity of time series which represents that system passing through different states.

*3.1 Kolmogorov complexity spectrum*

**Definition 1**. *The time series* $\{x_i\}$, $i = 1, 2, 3, 4, ..., N$ *we call normalized one after transformation* $x_i = (X_i - X_{min}) / (X_{max} - X_{min})$, *where* $\{X_i\}$ *is a time series obtained either by a measuring procedure or as an output from a physical model, while* $X_{max} = max\{X_i\}$ *and* $X_{min} = min\{X_i\}$.

**Remark**. From Def. 1 follows that all elements in time series $\{x_i\}$ lay in the interval $[0,1]$.

**Definition 2**. *The Kolmogorov complexity spectrum of time series* $\{x_i\}$ *we call the sequence* $\{c_i\}$, $i = 1, 2, 3, 4, ..., N$ *obtained by algorithm* $K$ *applied* $N$ *times on time series, where thresholds are all elements in* $\{x_i\}$.

**Remark**. The time series obtained either by a measuring procedure or as an output from a physical model, we transformed into a finite symbol string by comparison with series of thresholds $\{x_{t,i}\}$, $i = 1, 2, 3, 4, ..., N$, where each element is equal to the corresponding element in the considered time series $\{x_i\}$, $i = 1, 2, 3, 4, ..., N$, applying algorithm $K$. The original signal samples are converted into a 0-1 sequences $\{S_i^{(k)}\}$, $i = 1, 2, 3, 4, ..., N$, $k = 1, 2, 3, 4, ..., N$ defined by comparison with a threshold $x_{t,k}$,

$$S_i^{(k)} = \begin{cases} 0 & x_i < x_{t,k} \\ 1 & x_i \geq x_{t,k} \end{cases} . \qquad (4)$$

After we apply the algorithm $K$ (in our case the LZA) on each element of series $\{S_i^{(k)}\}$ we get the Kolmogorov complexity spectrum $\{c_i\}$, $i = 1, 2, 3, 4, ..., N$. This spectrum we introduce to explore the range of amplitudes in a time series representing a process, for which that process has highly enhanced stochastic components, i.e. highest complexity.

**Definition 3**. *The highest value* $K_{max}^C$ *in this series, i.e.* $K_{max}^C = max\{c_i\}$, *we call the Kolmogorov complexity spectrum highest value.*

We demonstrate meaning of measures we have introduced: (1) spectrum of the complexity $\{c_i\}$ and (2) Kolmogorov complexity spectrum highest value the spectrum $K_{max}^C$ (in further text KLM), using a time series $\{x_i\}$, through the following examples.

**Example 1**. In this example we demonstrate meaning of the complexity spectrum $\{c_i\}$. Here the time series $\{x_i\}$ we obtain by the instrument $\{M_\varepsilon, \tau\}$ with $M_\varepsilon = e^{-w\sigma}$, where $\sigma$ is random number uniformly distributed in the interval [0,1] while $w$ is the amplitude, which takes values in the interval [0,1]; $\{x_i\}$ is sampled every $\tau = 1$ time unit.

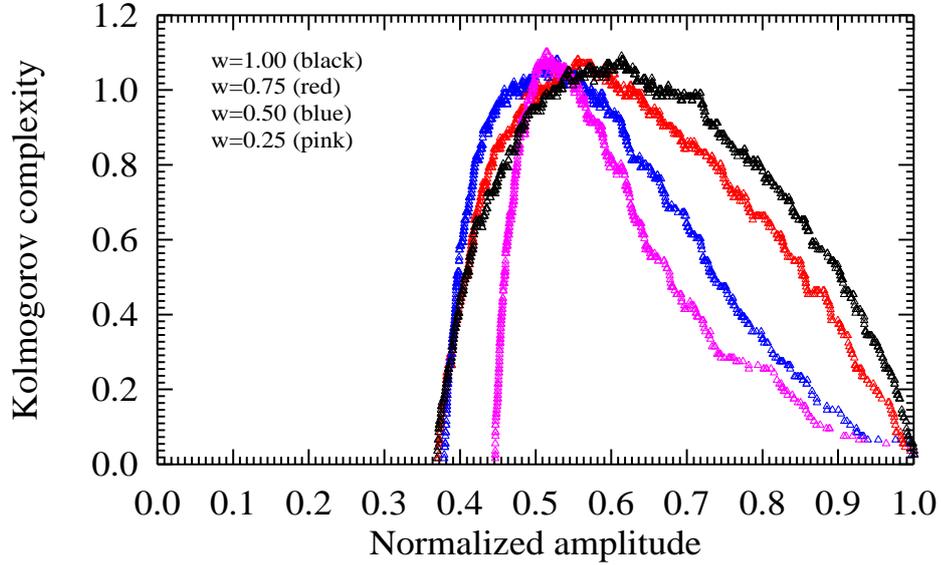

Fig. 2 The Kolmogorov complexity spectra $\{c_i\}$ of time series $\{x_i\}$ obtained by the instrument $\{M_\varepsilon, \tau\}$ with $M_\varepsilon = e^{-w\sigma}$, where $w$ is the amplitude factor, $\sigma$ is the random number uniformly distributed in the interval [0,1], and sampling $\tau = 1$ time unit.

Figure 2 shows the Kolmogorov complexity spectra for $w = 1.0$, 0.75, 0.50 and 0.25, respectively. From this figure we can see different spectra for different values of $w$. All of them are similar to curve in Fig. 1, which represents just one of many possible behaviors since different systems have different complexity vs. randomness plots since that there is no "universal" complexity-entropy relationship [13].

**Example 2**. To illustrate the sense of introducing the complexity measure $K^C_{max}$ we deal with, a time series $\{x_i\}$, $i = 1, 2, 3, 4, ..., 500$, which is generated by a generalized logistic map [25]. Mathematically, that map has the form

$$\Phi(x) = rx^p(1-x^p) \qquad (5)$$

where $r$ is a logistic parameter, $0 < r < 4$. This map expresses the exchange of biochemical substance between cells that is defined by a diffusion-like manner, where the parameter $p$ is the cell affinity. We have chosen this map because it is suitable to illustrate the meaning of $K^C_{max}$. We have calculated $K^C_{max}$ and $K^C$ complexities ($K^C$, calculated with threshold $x_t = \sum_{i=1}^{N} x_i / N$) for $p = 0.5$ and $p = 1$ ($0 < p \le 1$). In those computations, for each $r$ from 3.5 to 4.0 and $p$ with step 0.01, $10^3$ iterations were applied for an initial state, and then the first $10^2$ steps were abandoned. The results of computations are given in Fig. 3. From this figure it can be seen that in both cases $K^C$ carries less information about the complexity of the time series than $K_{max}^C$ does. Moreover, for $p = 0.5$ the $K^C$ is recognized only after $r > 3.9$ since it gives us average information about the complexity of the time series. In contrast to that $K_{max}^C$ carries the information about the highest complexity among all complexities in the spectrum. Therefore, this measure should be included in: (i) consideration how do we come to understand system's randomness and organization [5] and (ii) the complexity analysis of time series that an instrument provides. This measure can give deeper insights into these issues.

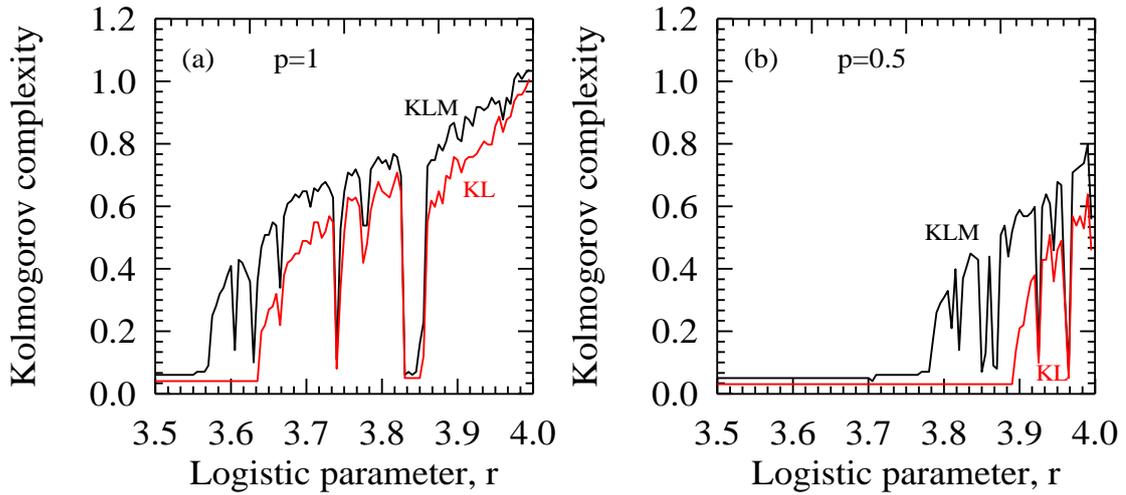

Fig. 3 The dependence on the logistic parameter $r$ of Kolmogorov complexity (KL, red) and the KLM (black) of time series generated by the generalized logistic equation $x_{n+1} = r x_n^p (1 - x_n^p)$ for (a) $p = 1$ and (b) $p = 0.5$.

In order to explore the dependence on the logistic parameter $r$ and cell affinity $p$ of (a) the KL and (b) the KLM, simulated by the generalized logistic Eq. (5). In those computations, for each $r$ from 0.0 to 4.0 and $p$ with step 0.01, $10^3$ iterations were applied for an initial state, and then the first $10^2$ steps were abandoned. Looking at Figs. 4a-4b, which depicts the KL and KLM complexities, respectively, we can see regions with different levels of complexity. Further inspection of figures points out that in the region of the KLM (Fig. 4b) its values are higher than for the KL ones (Fig. 4a). Apparently, that the KLM is better indicator about complexity the time series than commonly used the KL one. This is because the KL carries average information about a time series. In contrast to that the KLM carries the information about the highest complexity among all complexities in the Kolmogorov complexity spectrum.

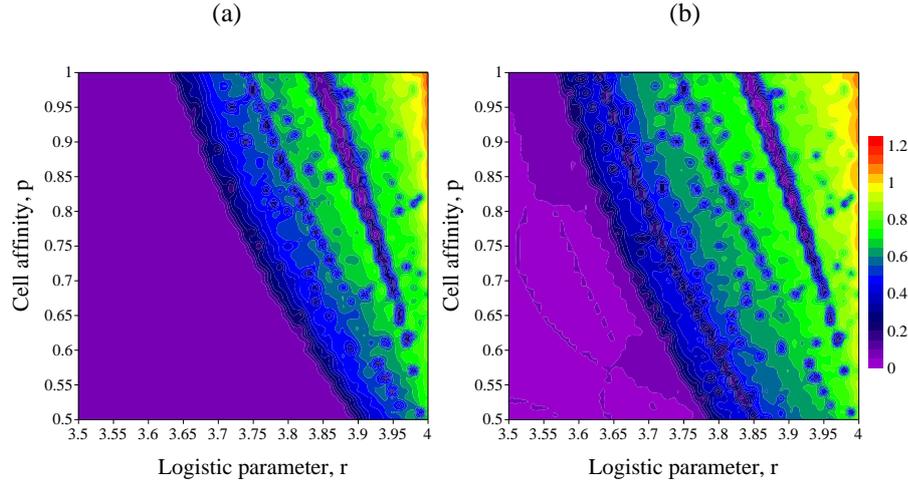

Fig. 4 The dependence on the logistic parameter $r$ and cell affinity $p$ of (a) Kolmogorov complexity (KL) and (b) Kolmogorov complexity spectrum highest value (KLM), simulated by the generalized logistic equation $\Phi(x) = rx^p(1-x^p)$.

*3.2 The overall Kolmogorov complexity measure*

We suggest a new measure based on the Kolmogorov complexity, which can be used for better understanding of physical complexity as well as the complexity of other systems, i.e. their time evolution and predictability. Before the elaboration we will shortly consider the term of the complexity from a perspective we would like to introduce this measure. Beyond any doubt this term means the possibility of the enrichment in the process of becoming, or less literarily - the possibility for a growth of structural complexity. A lot of papers regarding this issue have been offered during the last two decades. Among them we underline two contributions about complexity: (i) a comprehensive elaboration about this subject from different aspects (epistemological, mathematical as well as the physical), summariezed in [1] and (ii) a recent overview given by Crutchfield [5] who, in particular, have emphasized the difficulties in perception, which becomes more problematic when the phenomena of interest arise in systems that spontaneously organize.

To our mind the complexity of systems and its corresponding measures, could be considered in the following way. When a complex system is under observation only an active subject (a scientist, agent) that creates new communicative parameters of order allows the realisation of more complex information about a system that is connected with the idea of constructive chaos and chaos as a space of information. How can we gather the information about complexity (physical complexity in our case) expressed through some measure, particularly when the phenomena of interest arise in systems that spontaneously organize? This is the key question since the only available evidence about the nature of complex system is the agent's report written down in the form we call as time series. As we mentioned above according to Adami and Cerf [11], is to search for a measure of physical complexity as that it should closely correspond to our intuition with remark that it can consistently be defined within information theory. Finally, to close the agent context we emphasize a fact linked for its psychological structure that is not negligible. As cleverly observed by van der Pol and van der Mark [26], who noted that much of our positive reception depends on the fact whether our minds "ready" to confront with these intricacies or not. Further, they concluded that "When confronted by a phenomenon for which we are ill-prepared, we often simply fail to see it, although we may be looking directly at it".

Having in mind above attitudes we add to them the following consideration. First, for any complex physical system at any moment we can establish its entropy either through a measuring procedure or computation. That is only that can be written down in the agent's report at the fixed time (we refer as the "white window"). Second, between two successive agent's registration in the record there exits no information about complexity (except "that it should closely correspond to our intuition" [11] – nothing more and nothing less) – behind the window we refer as the "black window". Note that the complexity tell us how is complex transfer between two states with corresponding entropies we can measure or compute. The only we can do is to anticipate a measure of complexity, which will carry more information as much as possible.

The KL complexity as a measure is not able to recognize as distinct between time series with different amplitude variations and similar random components. On the other hand, the same feature could be also attached to the suggested KLM measure, although it gives more information about complexity, in a broader context, than the KL one does. Thus, when we convert a time series into a string then its complexity is hidden in the coding rules. For example, in the procedure of establishing a threshold for a criterion for coding some information about structure of the time series can be lost. However, from the spectrum of the Kolmogorov complexity $\{c_i\}$ of time series $\{x_i\}$ obtained by the instrument $\{M_\varepsilon, \tau\}$ we do not loose any information since we get $N$ fixed thresholds, each of them contributing to the dynamic of the system, and $N$ calculated complexities, i.e. corresponding spectrum (Fig. 2). The shape of this complexity curve depends on variability of time series amplitudes what cannot be captured by the KL and KLM. From that point of view the spectrum can be considered as a novel method in quantifying amplitude and complexity variations in time series. With introducing the way how the spectrum of complexity is computed, we increase the number of information "that is stored in a sequence about a particular environment", what is according to Adami [22] a definition of physical complexity of a sequence. The increase of number of information give us opportunity to have better access to insights of the system complexity since the physical or other complexities can be determined only by analyzing its correlation with the corresponding environment.

Figure 5 depicts the Kolmogorov based complexities of time series obtained by the instrument $\{M_\varepsilon, \tau\}$ as in subsection 3.1 in dependence on amplitude factor $w$. From this figure it is seen, that the KL values are very close for all amplitude factors. Similarly, it can be said for the KLM values. Apparently, neither the KL nor the KLM complexity is not able to discern between time series with different Kolmogorov spectra of complexity. From this reason we introduce *an overall Kolmogorov compelxity measure* $K_O^C$ (KLO in further text) defined as

$$K_O^C = \int_X K_s^C dx \qquad (6)$$

where $K_s^C$ is the spectrum of the Kolmogorov complexity, $dx$ is differential of the normalized amplitude, while $X$ is a domain of all normalized amplitudes, over which this integral takes values. Since $K_s^C$ is given as the sequence $\{c_i\}$, $i = 1, 2, 3, 4, ..., N$ (see Def. 2), it is calculated numerically as

$$K_O^C = c_1(x_2 - x_1) + \frac{1}{2}\sum_{i=1}^{N-1} c_i(x_{i+1} - x_{i-1}) + c_N(x_N - x_{N-1}) \qquad (7)$$

The $K_O^C$ takes value on the interval $(0, K_u)$, where according to Hu et al. [20] $K_u$ can take value up to $1.2$. This measure can make distinction between different time series having close values of the KL and KLM. It is clearly seen from Fig. 5 looking at the descending course of the KLO curve for different values of the amplitude factor $w$. Thus, only if information about the KLO is available we have made a reliable conclusion regarding the Kolmogorov complexity of time series.

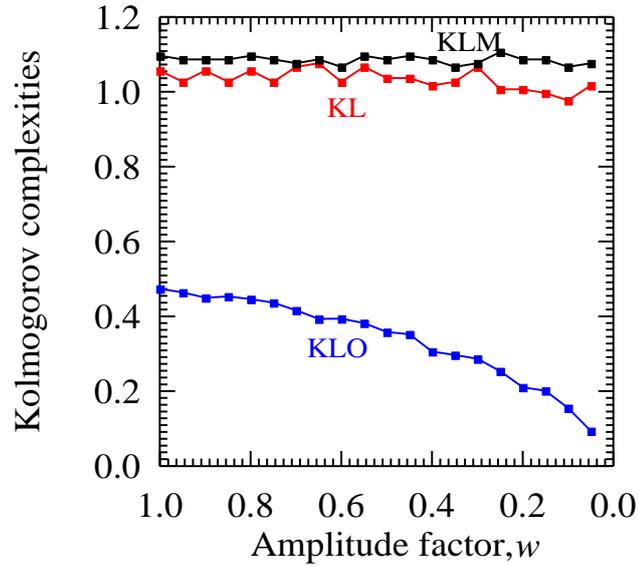

Fig. 5 The dependence on the amplitude factor $w$ of the Kolmogorov complexities (KL, KLM and KLO) of time series obtained by the instrument $\{M_\varepsilon, \tau\}$ as in subsection 3.1.

## 4. Application on different dynamical systems

In order to illustrate the performance of the introduced Kolmogorov based complexity measures, in real and modeling contexts of complex systems, we have used several modeling and natural records, similarly as it has been done by Zunino et al. [10].

| KL | KLM | KLO | Origin of time series | Time series |
|---|---|---|---|---|
|  |  |  | Modeled intra-cellular concentration in a multi-cell system | Different coupling parameters |
| 0.957 | 0.987 | 0.711 |  | c = 0.02 (c1) |
| 0.678 | 0.807 | 0.570 |  | c = 0.15 (c2) |
| 0.119 | 0.478 | 0.230 |  | c = 0.19 (c3) |
| 0.109 | 0.149 | 0.105 |  | c = 0.50 (c4) |
|  |  |  | River flow rate |  |
| 0.936 | 0.985 | 0.410 |  | Miljacka (Mil) |
| 0.936 | 0.973 | 0.406 |  | Bosnia (Bos) |

| | | | |
|---|---|---|---|
| | | | Long term precipitation amount |
| 1.152 | 1.152 | 0.511 | Bihac (Bi) |
| 1.097 | 1.097 | 0.558 | Mostar (Mo) |
| | | | Indoor $^{222}$Rn concentration |
| 0.891 | 0.934 | 0.302 | Year 2009 (Y09) |
| 0.832 | 0.832 | 0.323 | Year 2010 (Y10) |
| | | | UV radiation dose |
| 0.498 | 0.512 | 0.374 | Subotica (Su) |
| 0.497 | 0.527 | 0.382 | Zrenjanin (Zr) |
| | | | Stock prices |
| 0.978 | 1.013 | 0.218 | Imlek (IMLK) |
| 1.048 | 1.062 | 0.137 | Dean Food (DF) |

Table 1 Kolmogorov complexities (Kolmogorov complexity - KL; Kolmogorov complexity spectrum highest value - KLM and the overall Kolmogorov complexity measure - KLO) calculated for time series of different origin (the modeled intra-cellular concentration in a multi-cell system for different coupling parameters, the river flow rate, the long term precipitation amount, the indoor $^{222}$Rn concentration, the UV radiation dose and the stock prices).

*4.1 Intra-cellular concentration dynamics in a multi-cell system*

In the first application we choose to analyze the intra-cellular concentration dynamics in a multi-cell system represented by a ring of coupled cells (Fig. 6). In our approach, a cell moves locally in its environment without making long pathways. As a generalization of the two-cell system, according to Mihailovic et al [25,27], the dynamics of biochemical substance exchange in such a multi-cell system of $x_k^{(n)}$ cells, can be represented by the discrete nonlinear time-invariant dynamical system [28]:

$$\mathbf{x}^{(n+1)} = \mathbf{F}\left(\mathbf{x}^{(n)}\right) := C \Phi\left(\mathbf{x}^{(n)}\right) + (I - C) Z \Psi\left(\mathbf{x}^{(n)}\right), \tag{8}$$

where:
- $x_k^{(n)}$ is the concentration of the substance in $k$-th cell in a discrete time step $n$, $k = 1, 2, ..., K, n = 0, 1, 2, ..., N$ and $\mathbf{x}^{(n)} := \left[x_1^{(n)}\ x_2^{(n)}\ ...\ x_K^{(n)}\right]^T$ is the appropriate vector,
- $C := diag(c_1, c_2, ..., c_N)$ is the diagonal matrix of the coupling coefficients for each cell,
- $\Phi\left(x^{(n)}\right) := diag\left(\varphi(x_1^{(n)}), \varphi(x_2^{(n)}), ..., \varphi(x_K^{(n)})\right)$, is the diagonal matrix of intra-cellular behavior modeled by logistic map $\varphi: (0,1) \to (0,1)$, $\varphi(x) := r\, x(1-x)$,
- $\Psi\left(x^{(n)}\right) := diag\left(\left(x_2^{(n)}\right)^{p_1}, \left(x_3^{(n)}\right)^{p_2}, ..., \left(x_N^{(n)}\right)^{p_{K-1}}, \left(x_1^{(n)}\right)^{p_K}\right)$ is the diagonal matrix of the flow of the substance to each cell, where the all cell's affinities fulfill the constraint $p_1 + p_2 + ... p_K = 1$, and
- $Z \in \{0,1\}^{K \times K}$ is upper cyclic permutation matrix, i.e., $Z := \left[e_K\ e_1\ e_2\ ...\ e_{K-1}\right]$, where $e_1, e_2, ..., e_K$ are the standard basis vectors of $\square^N$.

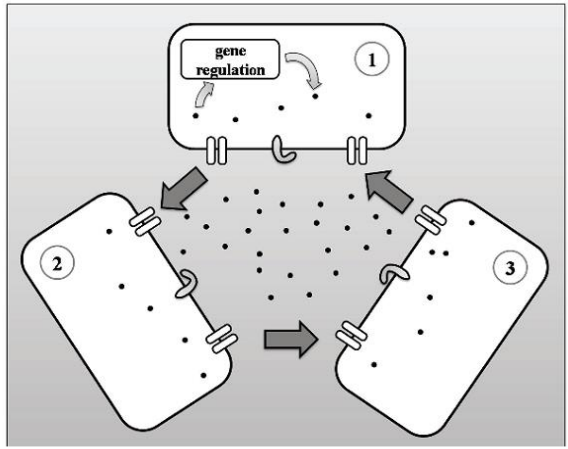

Fig. 6 Schematic diagram of a model of biochemical substance exchange in the system represented by a ring of coupled cells.

Simulations of biochemical substance exchange in the system represented by a ring of coupled $K = 3$ cells, given by Eq. (8), were performed with following values of parameters $r = 4$, $p_1 = p_2 = p_3 = 1/3$. The coupling parameter $c$ was taken to cover a broad range of coupling ranged from weak to strong (0.02, 0.15, 0.19 and 0.50) while the number of iterations was $N = 1000$. The results of simulations are depicted in Fig. 7. The curves, describing the Kolmogorov complexity spectra of time series of concentration, for different values of $c$, show significant difference in the complexities. Those difference are strongly correlated with the value of $c$, i.e. the complexity of the concentration dynamics is highest for the weakest coupling ($c = 0.02$) and it takes the lowest values for the strongest one ($c = 0.5$).

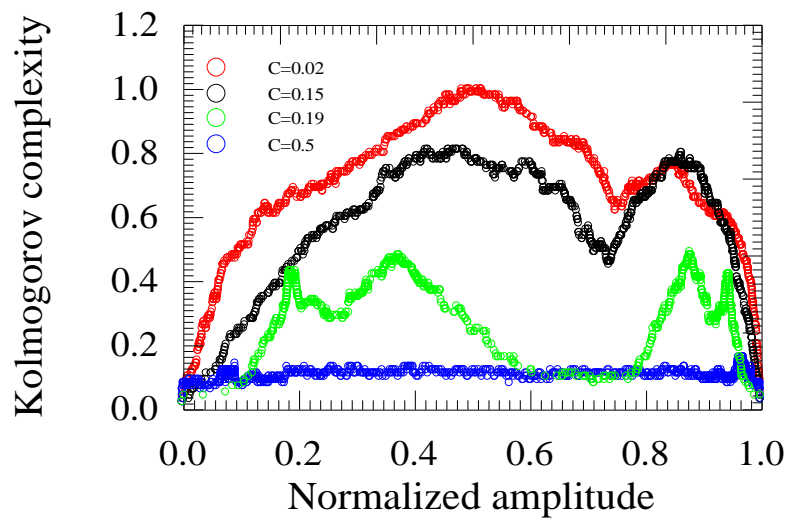

Fig. 7 The Kolmogorov complexity spectra for the normalized amplitude of the intracellular concentration obtained by the model of biochemical substance exchange in a system represented by a ring of coupled cells for four values of $c$.

The results of the KL, KLM and KLO calculations are given in Table 1. The order of the KL complexities (0.957, 0.678, 0.119, 0.109) for the $c1$, $c2$, $c3$ and $c4$ is pursued by the KLM complexities (0.987, 0.807, 0.478, 0.149) as well as with the KLO complexities (0.711, 0.570, 0.230, 0.105). Here, for these time series the hierarchy of all complexities is clearly enhanced. Therefore, in this case the KL measure carries enough information about complexity of this process.

*4.2 River flow dynamics*

Over the last decade controversial results have been obtained about the hypothetical chaotic nature of river flow dynamics [10,29,30]. For example, Zunino et al. [10] analyzed the streamflow data corresponding to the Grand River at Lansing (Michigan) trying to provide deeper insights regarding this issue, while Hajian and Sadegh Movahed [30] have used detrended cross-correlation analysis in order to investigate the influence of sun activity represented by sunspot numbers on river flow fluctuation as one of the climate indicators. The river flow fluctuations also have been analysed using the formalism of the fractal analysis [31]. We have analyzed the river flow time series corresponding to the Miljacka River and the Bosnia River (Bosnia and Herzegovina) for the period 1926–1990 on a monthly basis with $N = 780$ data points [32].

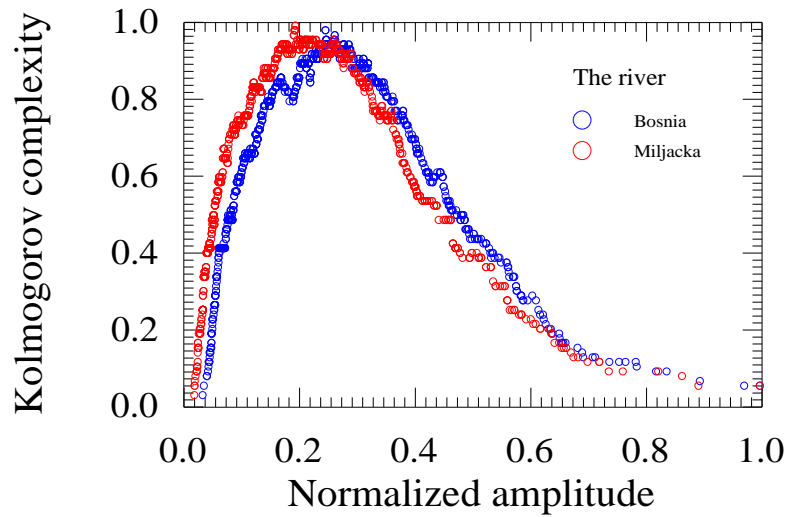

Fig. 8 The Kolmogorov complexity spectra for the normalized amplitude of the monthly flow rate time series of the Bosnia River and the Miljacka River (Bosnia and Herzegovina) for the period 1926–1990.

The results obtained are depicted in Fig. 8. The curves, describing the Kolmogorov complexity spectra for time series of flow rate of both rivers suggest that their flow dynamics are very similar with a pronounced presence of stochastic influence in these typically mountain rivers. The results of the KLM and KLO, given in the corresponding rows of Table 1, support this conclusion. Namely, the KLM values in both rivers are close (0.985 and 0.973 for Mil and Bos, respectively), while the KLO values practically are the same (0.410 for Mil and 0.406 for Bos). Thus, for this time series the KL (0.936 for both rivers) offers enough information about their complexity.

*4.3 Long term precipitation dynamics*

As a second geophysical application, we have analyzed the long term precipitation dynamics for two places with different geographical locations in Bosnia and Herzegovina. The pluviographic regime of one place is influenced by the vicinity of the Adriatic Sea (Mostar), while another one is surrounded by the mixed mountain and flat areas (Bihac). This example is chosen because to our knowledge this is the first time that the mentioned the KL complexity analysis is applied on a meteorological time series. The time series were updated for the time interval 1960-1984 having the length of $N = 300$ [33].

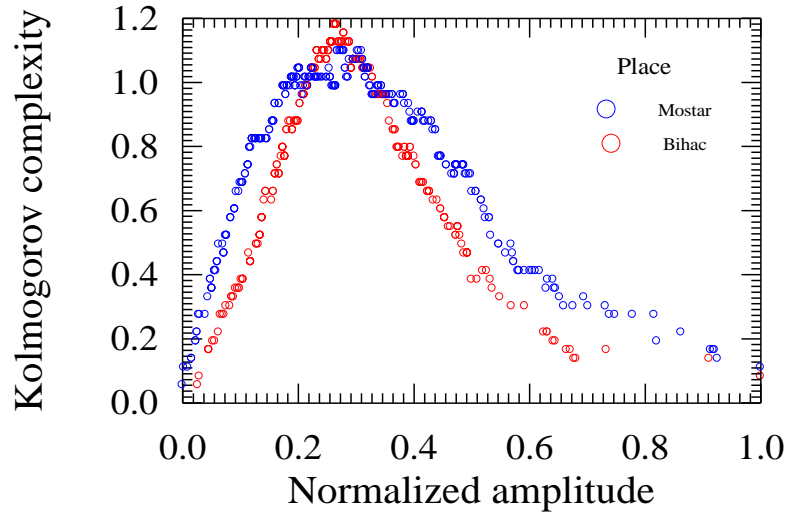

Fig. 9 The Kolmogorov complexity spectra for the normalized amplitude of the monthly precipitation amount time series for places Mostar and Bihac (Bosnia and Herzegovina) for the period 1960–1984.

The calculated values of complexities (the KL, KLM and KLO) are depicted in Fig. 9. The curves, describing the Kolmogorov complexity spectra for time series of both places suggest that their long term participation dynamics differs. The values of the KL are slightly different (1.097 - Bi and 1.152 - Mo) and for the KLM (1.097 - Bi and 1.152 - Mo) as it is presented in Table 1, while the difference in the KLO values is more pronounced (0.511 for Bi and 0.558 for Mo). Thus, the difference in the KL are negligible, but the difference in the KLO are larger. Therefore, we have to include this additional information about complexity that is not contained in KL and KLM. In conclusion we can say that the Mo time series is more complex than Bi time series, having larger variability of amplitudes as it is seen from the shape of its spectrum in Fig. 9.

*4.4 Indoor $^{222}$Rn concentration dynamics*

Data we needed for the analysis of the indoor $^{222}$Rn concentration dynamics we have obtained from the underground low-level laboratory in Belgrade. The special designed system for radon reduction, used in laboratory consists of three stage: (i) The active area of the laboratory is completely lined up with aluminium foil of 1 mm thickness, which is hermetically sealed with a silicon sealant to minimize the diffusion of radon from surrounding soil and concrete used for construction, (ii) the laboratory is continuously ventilated with fresh air, filtered through one rough filter for dust elimination followed by the battery of coarse and fine charcoal active filters and (iii) the parameters of the ventilation system are adjusted to give an overpressure of about 2 mbar over the atmospheric pressure. The radon monitor is used to investigate the temporal variations in the radon concentrations. For this type of short-term measurements the SN1029 radon monitor was used (manufactured by the Sun Nuclear Corporation). The device consists of two diffused junction photodiodes as a radon detector, and is furnished with sensors for temperature, barometric pressure and relative humidity. The user can set the measurement intervals from 10 min to 24 h. The radon monitor device records radon and atmospheric parameters readings every 10 min in the underground laboratory [34].

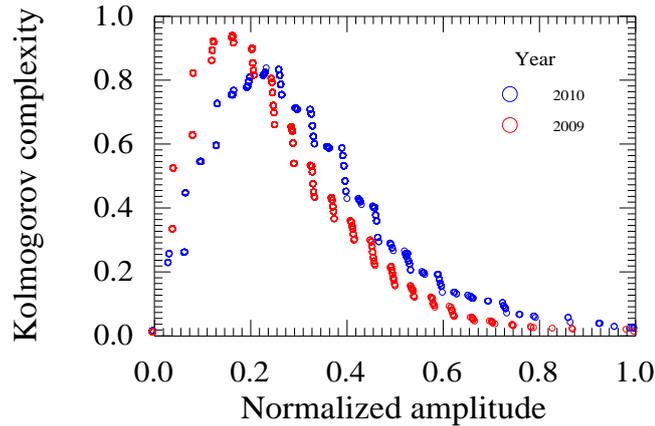

Fig. 10 The Kolmogorov complexity spectra for the normalized amplitude of the indoor $^{222}$Rn concentration time series created from data obtained measured in the Low-Background Laboratory for Nuclear Physics at the Institute of Physics in Belgrade (Serbia) for the period 1 January - 31 December 2009 and 2010.

We have analyzed the indoor $^{222}$Rn concentration time series corresponding to the data obtained from the Low-Background Laboratory for Nuclear Physics at the Institute of Physics in Belgrade (Serbia) for the period 1 January - 31 December 2009 and 2010 with $N = 4173$ data points for each time series. The results of the complexity analysis are given in Table 1, the calculated Kolmogorov complexity spectra are depicted in Fig. 10. The values of the KL are 0.891 (Y09) and 0.832 (Y10). The peaks in spectra show that the KLM values for the data sets Y09 (0.934) and Y10 (0.832) differs, and they are spaced on the scale of the normalized amplitudes. As we mentioned, this measure could be considered as a better indicator of the complexity in comparison with the KL, which is not always suitable measure of the complexity. In particular, this is enhanced in the case of asymmetrical distributions. Therefore, the time series for the Y09 data set is more complex than one for the Y10 data set. This points out only on the more emphasized presence of the stochastic component in indoor $^{222}$Rn concentration for the Y09 than for the Y10 data set. This can be attributed to the changed influence of indoor air temperature and relative humidity on $^{222}$Rn concentration [34]. However, in contrast to that the values of the KLO are different (Y09 - 0.302 and Y10 - 0.323). This indicates that the Y10 time series is more complex than Y09 one, having slightly larger variability of amplitudes. Therefore, the KLO could be considered as a suitable overall measure which gives deeper insights in the system dynamics.

*4.5 UV radiation dose dynamics*

To explore the UV-B radiation dose dynamics we have created the corresponding time series for Subotica (45.33° N, 19.85° E, 84 m a.s.l.) and Zrenjanin (45°24'N, 20°21'E, 80 m a.s.l.) in the Vojvodina region (Serbia). We have combined three sources because of the lack of measurement places for the UV radiation in this region. We have included: (i) measured values obtained by the broadband Yankee UVB-1 biometer, (ii) values computed by a parametric numerical model [35,36] and (iii) values computed by empirical formula based on linear correlation between the daily dose of the UV-B and the daily sum of the global solar radiation in MJ m$^{-2}$. The time series were updated for the time interval 1990-2007 having the length of $N = 6574$ [35].

The results obtained are shown in Fig. 11. The curves, describing the Kolmogorov complexity spectra for time series of both places suggest that their UV-B dose dynamics are quite similar without pronounced presence of stochastic components and with much lower values of the KL for both places (0.498 for Su and 0.497 for Zr) comparing to other time series (Table 1). There exists a slight difference in the KLM (0.512 for Su and 0.527 for Zr) and KLO (0.374 for Su and 0.382 for Zr). Although, the difference in the KL are practically the same, the difference in the KLO indicates that the Zr time series is slightly complex than Su one, having more pronounced variability of amplitudes.

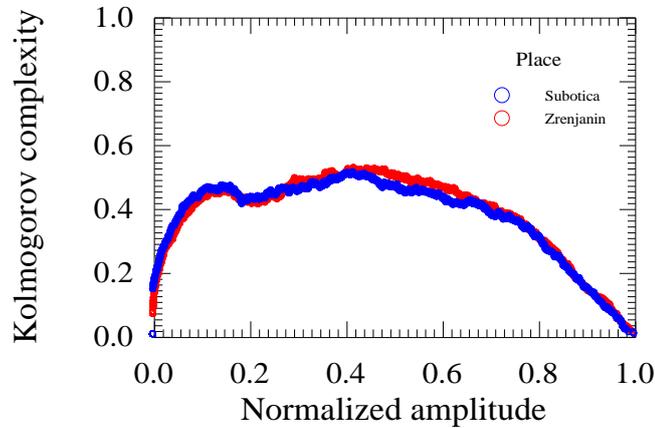

Fig. 11 The Kolmogorov complexity spectra for normalized amplitude of the daily UV-B dose time series for places Subotica and Zrenjanin (Vojvodina region, Serbia) for the period 1990–2007.

*4.7 Stock price dynamics*

One of the challenging problems in econophysics and financial econometrics is measuring market efficiency in terms of the pattern contained in price changes relative to patterns in random sequences. Market is efficient when price changes are unpredictable and random walk hypothesis is satisfied. This means that information is quickly incorporated in prices, eliminating possibility of market participants to profit from their information. In addition to the statistical approach in studying market efficiency also the approach based on information theory can be applied. Gulko [37] was the first who applied the concept of entropy in the analysis of financial series. Pincus and Kalman [38] used approximation entropy in studying market stability and Alvarez-Ramirez et al. [39] applied a multiscale entropy for measuring a time varying structure of market efficiency. Giglio et al. [40] applied the KL in order to rank stock exchanges and exchange rates. For the purpose of the comparison of the efficiency of stocks from developed and less developed markets we have chosen two time series. The first time series are daily closing stock prices of company Imlek (IMLK) from the Belgrade Stock Exchange. The company Imlek is a regional leader of dairy industry. The second time series is related to daily closing stock prices of Dean Foods (DF) from the New York Stock Exchange (NYSE), which is the largest processor and distributor of milk and other dairy products in the U.S. The sample period covers $N=1511$ trading days from 3 January 2011 to 30 December 2011.

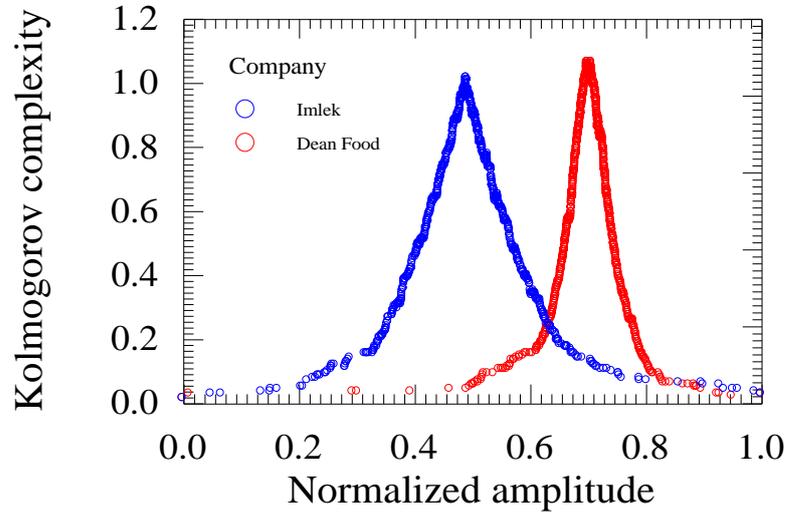

Fig. 12 The Kolmogorov complexity spectra for normalized daily returns of the companies Imlek from the Belgrade Stock Exchange and Dean Foods from the New York Stock Exchange.

The impact of non-stationarity, which characterizes both time series is reduced by converting the original series to returns, taking logarithm of the ratio of consecutive values of the series $r_i = log(p_i / p_{i-1})$, where $p_i$ are daily closing stock prices. The values of the KLM for both series of returns, given in the corresponding rows of Table 1, are greater than 1. That indicates their random behavior. As it is expected, the value of the KLM is lower for the stock from underdeveloped market. On the other hand, there is a larger difference between values of KLO i.e. between areas below the curves describing the Kolmogorov complexity spectra (Fig. 12). The curves may be compared with empirical density functions of normalized returns $r_i s = (r_i - min(r_i)) / (max(r_i) - min(r_i))$ (Fig. 13), estimated using Gaussian kernel [41].

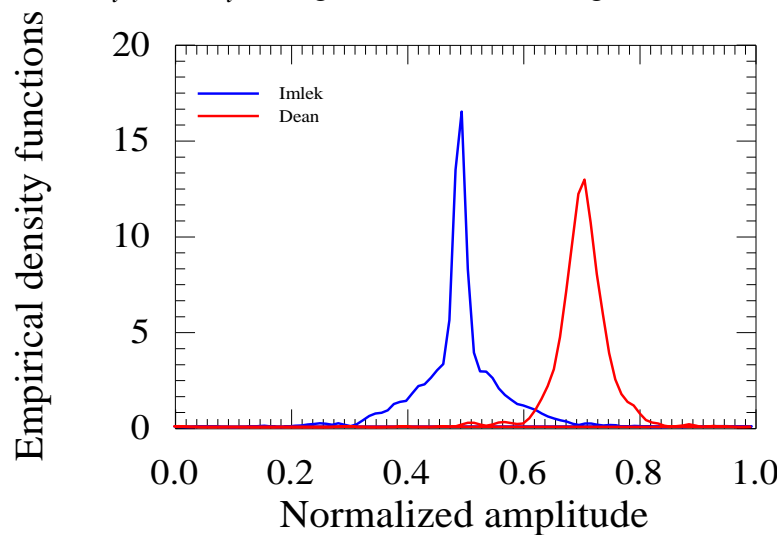

Fig. 13 Empirical density functions for normalized daily returns of the companies Imlek from the Belgrade Stock Exchange and Dean Foods from the New York Stock Exchange.

Both density curves are approximately symmetrical and have maximum values if normalized returns are close to medians of corresponding distributions $M_e^1 = 0.49$ and $M_e^1 = 0.70$. It can be noticed that each complexity curve reach a maximum value exactly for median values of normalized amplitude. The distributions of normalized returns (amplitudes) also differ in variability. The calculated values of standard deviations and coefficients of variation are: $sd(r_{1s})$= 0.08411851, $V(r_{1s})$= 17,8% for the IMLK time series and $sd(r_{2s})$= 0.05572103, $V(r_{2s})$= 7,94% for the DF time series. The difference in variability affects the shape of spectra of the Kolmogorov complexity curves. So it may be concluded that spectra of complexity gives the additional information about difference in amplitudes of time series that was not contained in the KL and KLM. The difference in spectra curves reflects on values of the KLO. The value of the KLO=0.218 for IMLK normalized returns is greater than the KLO=0.137 for DF.

## 5. Concluding remarks

In the present study we have proposed novel measures based on the Kolmogorov complexity for use in complex system behavior and time series analysis. We have considered Kolmogorov complexity, which is seen through its applicability in illuminating the physical as well as other complexities. Since the complexity of the system is hidden in the dynamics of the system we have introduced three novel measures based on the Kolmogorov complexity: the Kolmogorov complexity spectrum, the Kolmogorov complexity spectrum highest value and the overall Kolmogorov complexity. We apply the proposed measures on different time series originating from: the model output (the biochemical substance exchange in a multi-cell system), four different geophysical phenomena (river flow dynamics, long term precipitation dynamics, indoor $^{222}$Rn concentration dynamics and UV radiation dose dynamics) and economy (stock prices dynamics in the dairy industry). The result obtained show that we get deeper insights into the complexity if we use the introduced measures.

**Acknowledgements**

This paper was realized as a part of the project "Studying climate change and its influence on the environment: impacts, adaptation and mitigation" (43007) financed by the Ministry of Education and Science of the Republic of Serbia within the framework of integrated and interdisciplinary research for the period 2011-2014.